\documentclass[9pt,twocolumn,twoside]{osa-article}

\journal{oe}

\articletype{Research Article}

\usepackage[utf8]{inputenc}
\usepackage{graphicx}
\usepackage{float}
\usepackage{amsmath}
\usepackage{physics}

\usepackage[normalem]{ulem}
\newcommand{\stkout}[1]{\ifmmode\text{\sout{\ensuremath{#1}}}\else\sout{#1}\fi}

\begin{document}

\title{Observation of mode-mixing in the eigenmodes of an optical microcavity}

\author{C. Koks,\authormark{1,*} {M. P. van Exter,\authormark{1}}}

\address{\authormark{1}Huygens-Kamerlingh Onnes Laboratory, Leiden University,
P.O. Box 9504, 2300 RA Leiden, The Netherlands}

\email{\authormark{*}koks@physics.leidenuniv.nl} 



\begin{abstract}
We present a method to determine the complex coupling parameter of a two-coupled-modes system by directly measuring the coupled eigenmodes rather than their eigenvalues.
This method is useful because mode-mixing can be observed even if frequency shifts can not be measured.
It also allows to determine the complex coupling parameter, from which we conclude that the observed coupling is mainly conservative.
We observe mode-mixing in an optical microcavity, where the modes couple primarily at the mirror surface, as confirmed by AFM measurements.
The presented method is general and can be applied to other systems to measure mode coupling more accurately and to determine the nature of the coupling.
\end{abstract}

\section{Introduction}
Coupled harmonic oscillators occur in all fields of physics, including optics.
The coupling between harmonic oscillators or optical modes modifies the eigenmodes and shifts their eigenvalues.
We propose and demonstrate a method to directly observe the eigenmodes in an optical microcavity.
This is a sensitive method because it depends on the coupling amplitude instead of the coupling power; it thus allows one to also measure a small mode-mixing which is not visible in frequency shifts.
The complex amplitude also contains a phase, which reveals the nature of the coupling.

Optical microcavities are versatile and flexible tools to enhance the interaction between light and matter \cite{Vahala2003, Fushman2008}.
This enhancement, which is proportional to the cavity finesse divided by the mode area, can be controlled in an open microcavity \cite{Barbour2011,Greuter2014, Trichet2015,Potts2016, Wachter2019}. 
An open microcavity consists of two Distributed Bragg Reflectors (DBRs) with a tunable cavity length.
The radius of curvature of the DBR and the cavity length determine the mode size, and thereby the light-matter interaction.
Open microcavities can achieve similar Purcell factors as monolithic cavities \cite{Najer2019,Wang2019}.

Optical cavities support fundamental and higher-order transverse modes.
At certain cavity lengths, some modes become frequency degenerate and hence couple \cite{Klaassen2005, Benedikter2015, Benedikter2019, Trichet2018}.
The coupling of optical modes in analogous to two pendulums connected by a spring as depicted in Fig. \ref{fig: experimental setup}.
The modes of the pendulums hybridize and their eigenfrequencies shift.
Instead of measuring this frequency shift, we directly look at the motion of the pendulums and determine the mode-mixing amplitude from their positions.
The detection of the optical eigenmodes is more subtle because we measure intensities instead of electric fields.

In this paper, we report the direct observation of mode-mixing in far-field mode profiles and from this determine the complex coupling parameter.
This mode coupling is measured in a close to ideal system with an (almost) rotational-symmetric cavity.
The coupling is generated by a mismatch between the mode profile and mirror shape \cite{Kleckner2010} and by non-paraxial effects \cite{Luk1986, Erickson1975}.
The two modes that couple are identified and described by a generic model of two coupled harmonic oscillators.
We find the nature of the coupling to be conservative.

\section{Results}
\begin{figure}
    \centering
    \includegraphics[width=8.4cm]{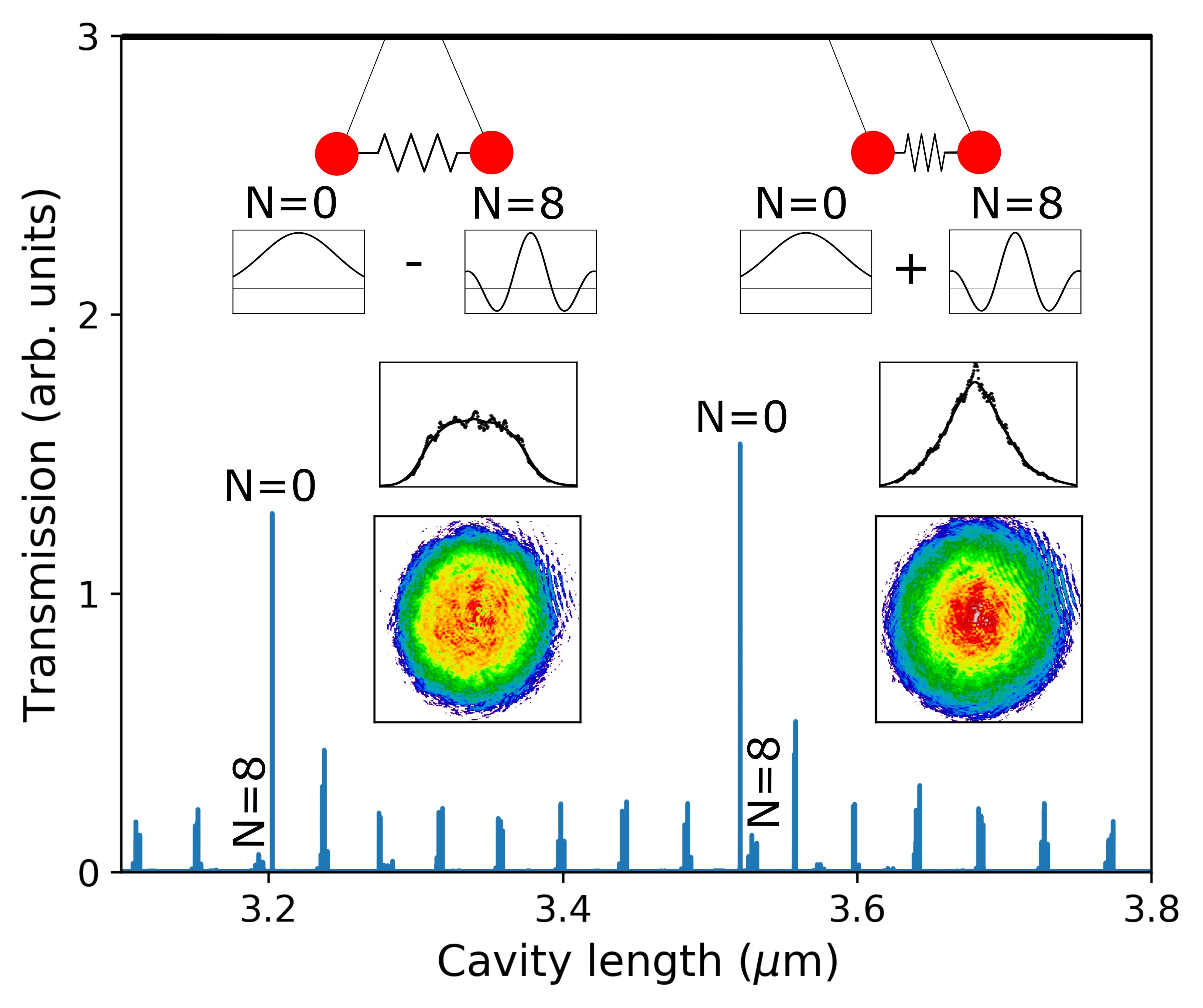}
    \caption{
    Transmission spectrum around coupling of the fundamental and $8^{\rm th}$ transverse mode.
    The insets show CCD images of the reshaped fundamental mode together with an averaged cross-section through the center. 
    The fundamental mode is coupled to the N=8 mode (insets only show central part), which can be in phase (right) or anti-phase (left).
    }
    \label{fig: experimental setup}
\end{figure}

Figure \ref{fig: experimental setup} shows a preview of our results in the form of power-normalized CCD images of two modes, close to frequency degeneracy.
These images show the center intensity is increased/decreased for positive/negative coupling.
A positive coupling reduces the effective mode area, theoretically up to a factor 2.
Mode coupling has been proposed as a means to increase the Purcell factor \cite{Podoliak2017}. 

The planar and concave mirrors of the microcavity are provided by Oxford HighQ and have a transmittance of $3.4(2)\times 10^{-5}$ and $1.1(1)\times10^{-4}$ at the wavelength $\lambda=633$ nm, close to the central wavelength of the DBR. 
The concave structures were fabricated with a focused ion beam \cite{Trichet2015} to produce craters with a radius $R\approx 24~\mu$m and an indentation depth $D\approx 0.6~\mu$m.

The mirrors are aligned to be parallel and almost in contact with each other.
The mirror distance is scanned over a total range of 15 $\mu$m using slip-stick motors and piezo stacks.
A HeNe laser ($\lambda=633$ nm) is focused into the cavity with an $f=8$ mm lens.
The light transmitted through the cavity is collected with another $f=8$ mm lens to measure the transmission spectrum and angular mode profiles. 

Figure \ref{fig: experimental setup} shows the transmission spectrum with sharp high-finesse peaks at particular cavity lengths.
These lengths are determined by the resonance condition, given below.
The fundamental modes, indicated by $N=0$ in Fig. \ref{fig: experimental setup}, are also measured with a CCD camera.
We use the angular mode profiles of the fundamental modes to demonstrate the mode coupling.

The paraxial eigenmodes in a rotational-symmetric cavity are Laguerre-Gaussian modes $\psi_{pl}$, labeled by their radial mode number $p$ and azimuthal mode number $l$ \cite{Siegman1986}.
The transverse mode number $N=2p+|l|$ and longitudinal mode number $q$ determine the resonant cavity lengths $L$ via the resonance condition $kL=q\pi+(N+1)\chi$, with wavevector $k=2\pi/\lambda$.
The Gouy phase $\chi=\sin^{-1}(\sqrt{(L+2L_D)/R})$, with modal penetration depth $L_D$ \cite{Koks2020}, quantifies the phase lag of the modes with respect to a plane wave.
The theoretically predicted opening angle of the fundamental mode is $\theta_0=\lambda/(\pi w_0)$, with mode waist $w_0$ and Rayleigh range $z_0=w_0^2 k/2=R \sin(2\chi)/2$.

\begin{figure} 
    \centering
    \includegraphics[width=8.4cm]{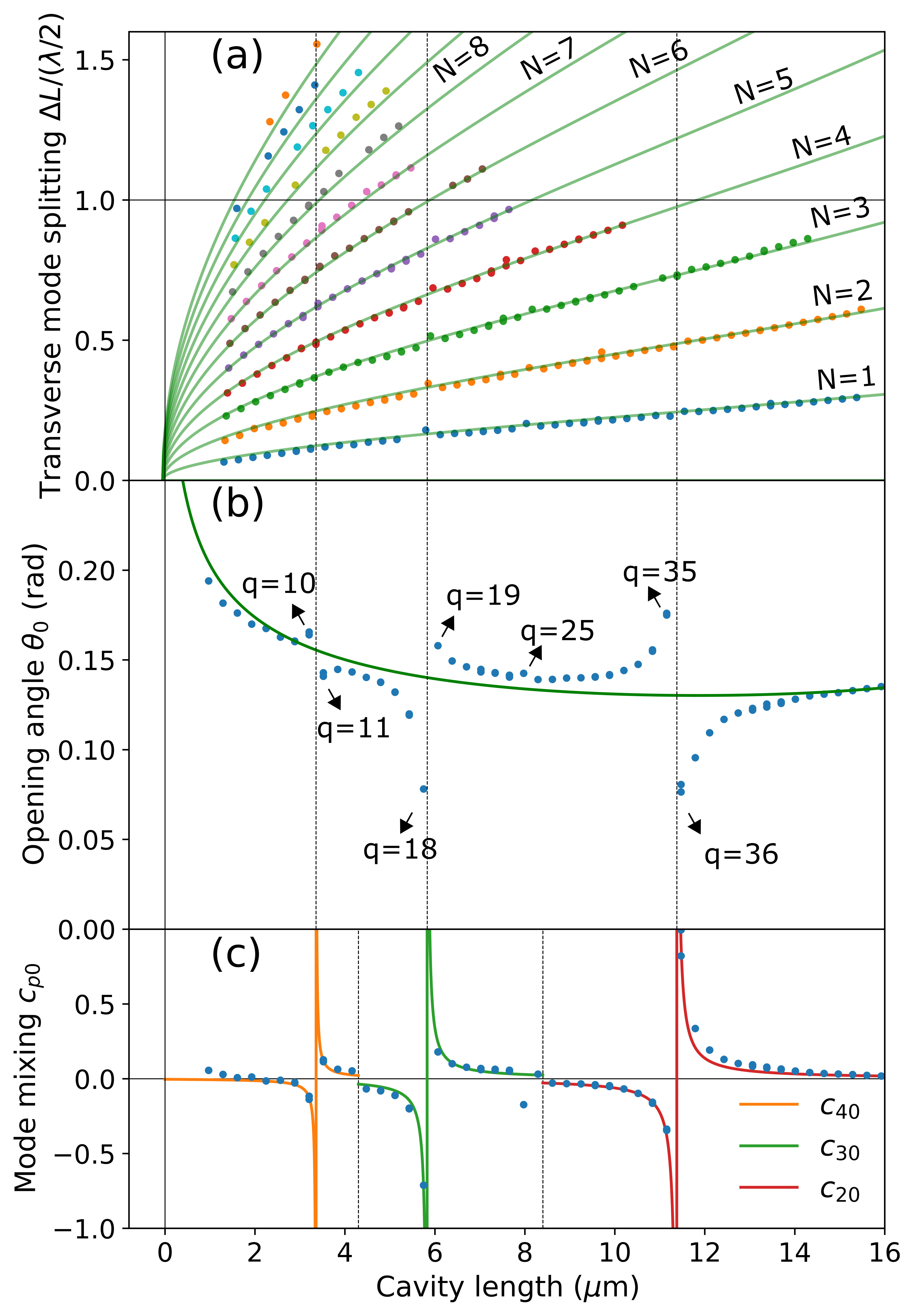}
    \caption{
    (a) Transverse mode splitting versus mirror position for all visible transverse modes.
    (b) The Gaussian opening angle $\theta_0$ of the fundamental mode obtained from CCD images. The green line shows the theory for the uncoupled system.
    (c) Mode-mixing ratio $c_{p0}$ of modes $p=4,3$, and $2$ into the fundamental mode. 
    }
    \label{fig:transverse and opening angle}
\end{figure}

Figure \ref{fig:transverse and opening angle}a shows the measured transverse mode splitting as a function of cavity length.
We plot the difference in resonant cavity length $\Delta L$ between the fundamental ($N=0$) and the $N^\text{th}$ order mode to find the Gouy phase using the relation $\Delta L/(\lambda /2) =N \chi/\pi$. 
A fit of the data from the $N=1-5$ modes yields a radius $R=23.8(2)~\mu$m and modal penetration depth $L_D=0.03(2)~\mu$m.
The first visible longitudinal mode is $q=3$ because the smallest cavity length is at least as large as the indentation depth $D=0.64(3) ~\mu$m.

Figure \ref{fig:transverse and opening angle}b shows the opening angles $\theta_0$ of the fundamental modes.
Each point in the graph corresponds to a Gaussian fit of a CCD image.
The mode profile is obtained by imaging the far-field, rather than the near-field, and is hence less sensitive to imaging aberration.

The general trend of the Gaussian fits in figure \ref{fig:transverse and opening angle}b follows the theoretical prediction (green curve), which is based on the parameters extracted from \ref{fig:transverse and opening angle}a and contains no fit parameters.
The measured opening angle, however, strongly deviates from theory around three cavity lengths, indicated by black vertical lines in Figs. \ref{fig:transverse and opening angle}ab.
At these cavity lengths, the mode profile deviates from a Gaussian and exhibits features of mode-mixing.
This occurs when the even transverse modes $N=8,6,4$ cross the line $\Delta L/(\lambda/2)=1$ (see Fig. \ref{fig:transverse and opening angle}a) and hence become frequency degenerate with the fundamental mode. 
The dominant mixing with even modes suggests a rotational-symmetric coupling effect.
The modest deviation at the point indicated by $q=25$ also indicates some mixing with $N=5$ modes, but this mixing is significantly smaller. 
Modest deviations are also observed for points at the beginning, where the mode waist is somewhat smaller.
The effective radius of curvature is larger for these small modes, as confirmed by atomic force microscopy (AFM) measurements (see supplemental document).

\begin{figure}
    \centering
    \includegraphics[width=8.4cm]{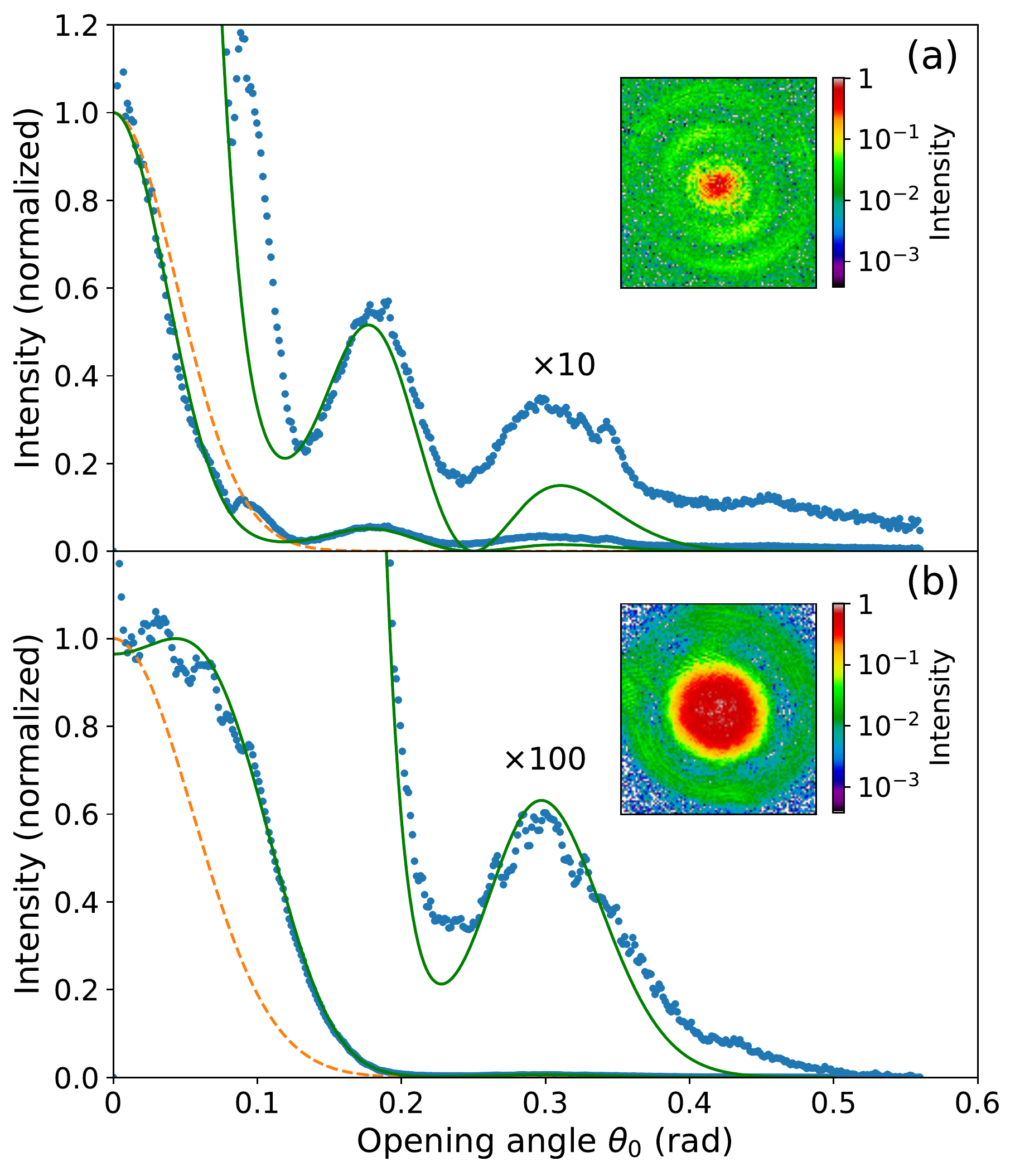}
    \caption{Rotation-averaged intensity of angular mode profile for (a) q=18 and (b) q=19. The dots show the experimental results from the 2D intensity profiles shown as insets. The orange dashed curves show the uncoupled Gaussian shapes. The smooth green curves show the fitted two-mode shapes with fit parameters (a) $c_{30}=-0.70$ and (b) $c_{30}=0.19$.}
    \label{fig:p=3 coupling}
\end{figure}

Figure \ref{fig:p=3 coupling} explains how the mode mixing in Fig. \ref{fig:transverse and opening angle}c is quantified. Two angular mode profiles are shown, which are observed at cavity lengths corresponding to $q=18$ and $q=19$.
These profiles show rings at larger angles, which indicates that a Gaussian fit with only the fundamental mode $\psi_{00}$ no longer suffices and a two-mode fit is required.
The $l=0$ mode dominates the mixing, due to the rotational symmetry of the cavity, such that $N=2 p$ for the coupled modes.
For the two-mode fit, we use the amplitude profile $\psi=(\alpha\psi_{00}+(-1)^p \beta \psi_{p0})$, with a complex mixing ratio $c_{p0}=\beta/\alpha$ and real-valued field-profiles $\psi_{00}$ and $\psi_{p0}$.
The phase lag between the fundamental and the higher-order mode from the near-field to the far-field is incorporated in the factor $(-1)^p$, such that the amplitude of the coupling constant relates directly to the field profiles at the flat mirror.
If $c_{p0}$ is positive and real-valued, the central part of the fields interfere constructively at the flat mirror and destructively at the curved mirror.
We take $c_{p0}$ to be real-valued because the mode profiles show strong interference effects.
The imaginary part of $c_{p0}$ only results in an incoherent admixture of typically 4\% in power, which we assign to residual background effects.
We fit the intensity profiles in Fig. \ref{fig:p=3 coupling} with a two-mode fit $|\psi_{00} - c_{3 0} \psi_{3 0}|^2$. 
This yields mixing ratios $c_{30} = -0.70$ and $0.19$ for modes $q=18$ and $19$, respectively.

Figure \ref{fig:transverse and opening angle}c shows the mixing ratio $c_{p0}$ for all cavity lengths. 
To obtain these data, we fit all CCD images with two-mode fits rather than Gaussian fits.
The theoretical opening angle $\theta_0$ from Fig. \ref{fig:transverse and opening angle}b is used to describe the uncoupled modes.
Three regions in Fig. \ref{fig:transverse and opening angle}c are identified in which the fundamental mode $\psi_{00}$ couples either with $\psi_{40}$, $\psi_{30}$ or $\psi_{20}$. 
Substantial mode-mixing is observed around the frequency degenerate points.
The mode-mixing with $\psi_{40}$ ($N=8$) shown in Fig. \ref{fig: experimental setup} is slightly weaker than the mixing with $\psi_{30}$ shown in figure \ref{fig:p=3 coupling}.
The mixing with $\psi_{20}$ ($N=4$) around $11.3~\mu$m shows signatures of more than 2 modes mixing in the CCD images (figure not shown). 

Coupled cavity modes behave like coupled harmonic oscillators (see supplemental document).
Two modes, continuously excited by an input field through a mirror with transmission $t$, reach an equilibrium described by
\begin{equation}
\label{eq:coupled modes equation}
    \begin{pmatrix} i(\varphi_a-\varphi) + \gamma_{a} & -M_{ab} \\ -M_{ba} & i(\varphi_b-\varphi) + \gamma_{b} \end{pmatrix}
    \begin{pmatrix} \alpha \\ \beta \end{pmatrix}_{c}
    =t \begin{pmatrix} \alpha \\ \beta \end{pmatrix}_{in}
\end{equation}
where the parameters in the matrix are dimensionless and describe variations per roundtrip.
The roundtrip phase of a plane wave is given by $\varphi=2 k L$, and the roundtrip phase of the $N^{\rm th}$ transverse eigenmode is $\varphi_{a/b}= 2\pi q+ 2(N+1)\chi(L)$.
Note that each uncoupled mode is resonant if $\varphi=\varphi_{a/b}$. 
The roundtrip losses $\gamma_{a/b}$ determine the finesse of the uncoupled modes via $F=\pi/\gamma$.

The two modes couple at the concave mirror, where a mismatch between the mirror shape and the wavefront causes light to scatter from mode $\psi_a$ into mode $\psi_b$ and vice versa. 
This coupling is quantified by a dimensionless coupling parameter $M_{ab}=\mel{\psi_a}{2 i k \Delta z}{\psi_b}$.
The mirror-mode mismatch $\Delta z$ has two contributions. 
First, it contains the deviations of the mirror from a paraboloid.
Second, it contains non-paraxial effects, which cause the wavefront to deviate from a paraboloid (see supplemental document).
The first contribution is dominant in our microcavities.

The coupled-harmonic-oscillator model is used to fit the data in Fig. \ref{fig:transverse and opening angle}c.
The solution to Eq. (\ref{eq:coupled modes equation}) predicts a mixing ratio $c_{ab}=\frac{M_{ba}}{i\Delta\varphi+\gamma_{ab}}$ for small enough coupling parameters (see supplemental document).
In our measurement, the detuning $\Delta\varphi=\varphi_b-\varphi_a$ is typically much larger than  $\gamma_{ab}=\gamma_b-\gamma_a$, so that the latter can safely be neglected.
A fit of the observed mixing ratios $c_{p0}(\Delta \varphi)$ in the three regions gives $M_{40}=0.018(2) i$, $M_{30}=0.034(3) i$ and $M_{20}=0.029(1) i$. 

All three values of $M_{p0}$ are purely imaginary with a positive imaginary part.
This is experimentally evidenced by the strong interference effects in Fig. \ref{fig:p=3 coupling}, which correspond to real-valued mixing ratios $c_{ab}$.
The coupling must hence be due to a wavefront mismatch at the curved mirror, and not due to clipping losses at the edges of the mirror.
The positive sign for all three couplings suggests that the wavefront mismatch dominantly occurs at the center of the curved mirror (see supplemental document).

To find the precise origin of the coupling, we have measured the shape of the concave mirror with AFM imaging.
We find a rotational-symmetric defect, which elevates the central part of the concave mirror by 0.08(2) $\mu$m with respect to the ideal parabolic shape with radius of curvature $R=23.8~\mu$m  (see supplemental document).
The coupling parameters $M_{p0}$ calculated from this mirror height profile are $M_{40}=0.015 i$, $M_{30}=0.049 i$ and $M_{20}=0.042 i$. 
The  coupling $M_{40}$ agrees reasonably well with the optical data, but the AFM-based estimations of $M_{30}$ and $M_{20}$ are a factor 1.4 larger than the optical measurements. 
This discrepancy can partially be assigned to a non-paraxial correction, which reduces the calculated mode coupling for $p=2$ by $\Delta M_{20}\approx0.004 i$, and to optically transparent height defects on the micromirror which only shown up in AFM measurements.

Measurements on different cavities have shown the same mode-coupling effects (see supplemental document).
The magnitude of the coupling is similar to that of the cavity presented here.
Also the sign is similar, which suggests that the effect that causes the coupling is similar.
This shows that the effect is general and occurs in different systems.

\section{Conclusion}
In summary, we have accurately measured the intensity profiles and opening angles of the fundamental mode in a microcavity as a function of cavity length. 
The general trend is as expected, but strong deviations were observed around three cavity lengths, where the fundamental mode couples with different higher-order modes.
The coupling is conservative and is attributed to a mismatch between the mirror shape and the wavefront.
The measured mode-mixing ratios near the frequency-degenerate points are substantial. 
This can potentially reduce the mode area and increase the Purcell factor, theoretically up to a factor 2 \cite{Podoliak2017}.

Rather than measuring an avoided crossing in the frequency spectrum, we observe the mode coupling directly in the far-field mode profile. 
This is a sensitive and powerful method, which directly yields the complex mixing ratio $c_{ab}$ from which the complex coupling parameter $M_{ab}$ is derived.
Frequency shifts or dips in finesse from mode coupling are more difficult to measure, since these scale quadratically instead of linearly with the coupling parameter.
The amplitude and phase of the coupling parameter provide information about the nature of the coupling. 

\begin{backmatter}

\bmsection{Acknowledgments} We like to acknowledge M.J.A. de Dood and X. Chen for fruitful scientific discussions and help with preparing the manuscript.

\bmsection{Disclosures} The authors declare no conflicts of interest.

\bmsection{Supplemental document} See Supplement 1 for supporting content. 

\end{backmatter}

\bibliography{main.bbl}

\end{document}


\title{Observation of mode-mixing in the eigenmodes of an optical microcavity: supplemental document}
\author{}

\begin{abstract} 
This supplement contains four sections: (i) it presents equations for the roundtrip operator $M$, (ii) it links the evolution of the intracavity field to the dynamics of the coupled cavity modes, (iii) it presents the analysis of AFM measurements of the mirror shape and links this to the shape-induced mode coupling, and (iv) it presents measurements on another cavity to demonstrate the generic behaviour of mode coupling. 
\end{abstract}

\maketitle

\section{Roundtrip operator $M$}

The evolution of the optical field in a cavity can be described by the roundtrip operator $M$, which transforms the forward-propagating intra-cavity field $| \psi^+ \rangle$ into $M | \psi^+ \rangle$ after a roundtrip \cite{Kleckner2010}. 
If we expand the field $| \psi^+ \rangle = \sum_i c_i | \psi_i^+ \rangle$ in a set of orthogonal basis states $|\psi^+_i \rangle$, we can write $| \psi^+ \rangle$ as a vector and $M$ as a matrix. 
Finding the resonance conditions and eigenmodes of the cavity now boils down to finding the eigenvalues and eigenvectors of the roundtrip matrix. 

We consider a plano-concave cavity with a large smooth planar mirror and a smaller concave mirror.
The reflection at the large planar mirror results in eigenmodes with a flat wavefront at this mirror and forward and backward propagating fields $| \psi^+ \rangle$ and $| \psi^- \rangle$ that are mirror imaged copies of each other.
This symmetry allows one to write the round-trip operator as a product $M=PB$, where $P$ describes the roundtrip propagation and $B$ describes the reflection from the concave mirror.

Kleckner et al. \cite{Kleckner2010} present an equation for the elements $B_{s,t}$ of the reflection matrix, which couple a forward-propagating field $| \psi_t^+ \rangle$ into a backward-propagating field $| \psi_s^- \rangle$. 
Unfortunately, their Eq. (5) is the complex conjugate of the correct expression
\begin{equation}
\label{eq:Kleckner}
    B_{s,t} = \langle \psi^-_s | B |  \psi^+_t \rangle = \int_0^{r_{max}} \int_0^{2\pi} r dr d \theta \,\, \psi_s^{-*} \psi_t^{+} e^{-2ikz_{\rm mirror}} \,,
\end{equation}
where the integral, in polar coordinates $r$ and $\theta$, runs over the finite extent of the mirror.
This equation simply states that the reflected field is a copy of the input field with a modified phase front and size, assuming perfect mirror reflection. 
The function $z_{\rm mirror}(r,\theta)$ describes the shape of the concave mirror, where $z>0$ points towards the planar mirror such that $z_{\rm mirror}(r,\theta) = r^2/(2R)$ for a paraboloidal concave mirror with radius $R > 0$.

It is instructive to discuss some properties of the roundtrip operator $M = r_1 r_2 \exp(i2kL) B$, where $r_1$ and $r_2$ are the reflection coefficients of the mirrors. 
The associated roundtrip matrix is symmetric, i.e. $M_{ij} = M_{ji}$, even if the roundtrip phase delays of the two modes - hidden in the phases of $\psi^\pm$ - differs.
The roundtrip operator is unitary and the system is lossless (energy conserving) when mirror B is large ($r_{max} \rightarrow \infty$) and when both mirrors are perfect reflectors ($r_1=r_2=1$).
Scattering is described by the off-diagonal elements of $M$, which quantify the (energy-conserving) scattering between the different modes per roundtrip.

The matrix elements of $M$ and $B$ depend on the choice of basis states. 
For our geometry, the obvious choice is the set of Laguerre-Gaussian (LG) modes with waist $w_0$ at the plane mirror and wavefront radius $R_{\rm mode} = L + z_0^2/L = R$ at the concave mirror (Rayleigh range $z_0 = \frac{1}{2}kw_0^2$). 
The modes in this set are
\begin{equation}
    \psi_{p\ell}^\pm (r,\theta) = |\psi_{p\ell}(r)| \, \exp[\pm i(\frac{kr^2}{2R}-\chi_{p\ell})+i\ell\theta] \,,
\end{equation}
at $z=L$, where $p$ and $\ell$ are the radial and angular quantum number that together label the modes.
The Gouy phase $\chi_{p\ell} = (2p+|\ell|+1) \chi_0$, with $\chi_0 = \arcsin{\sqrt{L/R}} = \arctan(L/z_0)$, is the single-pass phase lag that LG-modes acquire with respect to a plane wave.
The amplitude function $|\psi_{p\ell}(r)| = C_{p\ell} x^{|\ell|/2} L_p^\ell(x) \exp(-x/2)$, with $x=2r^2/w^2$ and normalization constant $C_{p\ell}$, describes the usual LG-mode profiles.
The angular momentum $\ell$ does not change in a coordinate system with fixed orientation.
\\
Next, we will calculate the roundtrip matrix for a rotational-symmetric cavity, where $\ell$ is conserved, with $r_1 = r_2 \approx 1$. 
The on-diagonal elements are
\begin{eqnarray}
\label{eq:Maa}
    M_{aa} & = & \exp(2ikL-2i\chi_a) \int_0^{r_{max}} \pi d r^2 \, |\psi_a^+(r)|^2 \, e^{-2ik\Delta z(r)} \nonumber \\
    & \approx & \exp(2ikL-2i\chi_a-2i\Delta \chi_a) \times  \nonumber \\
    & & \left(1 - \int_0^{r_{max}} \pi d r^2 \, 2k^2[\Delta z(r)]^2 \, |\psi_a^+(r)|^2  
    - \int_{r_{max}}^\infty \pi d r^2 \, |\psi_a^+(r)|^2 \right) \,,
\end{eqnarray}
where $\Delta z = z_{\rm mirror} - z_{\rm mode}$. 
The first equation combines Eq. (\ref{eq:Kleckner}) with $M = \exp(i2kL) B$ and with the Gouy phase lag $\chi_a$ of the paraxial modes.
The second equation is based on the Taylor expansion $\exp(-2ik\Delta z) \approx 1 - 2ik\Delta z - 2k^2[\Delta z]^2$.
The imaginary-valued term in this expansion results in a shift $\Delta \chi_a$ of the resonance condition.
The quadratic term in this expansion quantifies the scattering loss induced by height variations.  
The final term in Eq. (\ref{eq:Maa}) quantifies the clipping loss at a mirror of radius $r_{max}$.
Both losses are amplitude losses.
The middle term in Eq. (\ref{eq:Maa}) thus yields the familiar expression $(4\pi\sigma/\lambda)^2$ for the intensity loss, where $\sigma^2 = \langle \Delta z^2 \rangle$ is the mean-square height difference between mirror and phase front.

The coupling between modes $a$ and $b$ is described by the off-diagonal element 
\begin{eqnarray}
\label{eq:Mab}
    M_{ab} = M_{ba} =  \int_0^{r_{max}} \pi d r^2 \, |\psi_a^+(r) \psi_b^+(r)| e^{-2ik\Delta z(r)} \approx \hspace{34mm}  \nonumber \\
   \hspace{10mm} - \int_{r_{max}}^\infty \pi d r^2 \, |\psi_a^+(r) \psi_b^+(r)| - 2ik \int_0^{r_{max}} \pi d r^2 \, \Delta z(r) \, |\psi_a^+(r) \psi_b^+(r)| \,, 
\end{eqnarray}
where $\Delta z = z_{\rm mirror} - (z_{\rm a}+z_{\rm b})/2$ and where we again assumed $k\Delta z \ll 1$.
Modes with different angular momentum $\ell$ do not coupling in a rotational-symmetric cavity. 

We distinguish between two forms of coupling: dissipative and conservative.
Dissipate coupling is associated with energy loss and, in our system, originates from clipping at the mirror. 
It is described by the first term in Eq. (\ref{eq:Mab}) and produces a real-valued $M_{ab}$.
Conservative coupling is energy conserving and, in our system, originates from deformations of the wavefront upon reflection. 
It is described by the second term in Eq. (\ref{eq:Mab}) and produces an imaginary-valued $M_{ab}$.

Equations (\ref{eq:Maa}) and (\ref{eq:Mab}) show that the coupling $M_{ab}$ is linear in $k\Delta z$, whereas the scattering loss in $M_{aa}$ is quadratic in $k\Delta z$.
The relation between these two quantities is as follows.  
The relative intensity loss by scattering from mode $a$ to mode $b$ is $|M_{ba}|^2$ per roundtrip, making the scattering loss to all discrete modes equal to $\sum_b |M_{ba}|^2$.
This sum must be equal to the total scattering loss, $1 - M_{aa}^2$, from mode $a$. 
This relation is a consequence of the conservative character of the operator $M$ in the large-mirror-limit ($r_{max} \rightarrow \infty$).

For our cavity with large concave mirror, dissipative coupling is expected to be small and $M_{ab}$ will thus be (mainly) imaginary-valued and described by the second term in Eq. (\ref{eq:Mab}). 
To support this statement we calculate the expected clipping losses for three modes at a typical cavity length $L=L_3=5.8~\mu$m, where the $p=0$ and $p=3$ modes are frequency degenerate.
For this calculation we also need the dimensions of the concave mirror (radius of curvature $R=23.8~\mu$m, radius $r_{max} = 5.0~\mu$m) and the calculated size of the fundamental mode $w_z = 1.65~\mu$m, with $I(r) \propto \exp(-2r^2/w_z^2)$, at this mirror. 
The calculated (amplitude) clipping loss of the fundamental ($p=\ell=0$) mode is order $10^{-8}$ and hence negligible. 
The calculated (amplitude) clipping loss of the ($p=3, \ell=0$) mode is $6\times10^{-3}$, which limits its modal finesse to $F \approx 250$, to be compared with an observed finesse of $F \approx 5000$ for the fundamental mode. 
But even a loss $\gamma_b = 6 \times 10^{-3}$ is relatively small as compared to a typical detuning $\varphi_a - \varphi_b$ and can hence be neglected (see below).

We end this section with a disclaimer.
The description presented above is incomplete because the LG-modes are only the paraxial solutions of the full propagation. 
Three non-paraxial corrections should be added to the propagator $P$ when the opening angles are too large, i.e. when the modes are compact ($w_0 \approx \lambda$) \cite{Cullen1979,Erickson1975,Yu1983,Luk1986}. 
These non-paraxial corrections lead to small modifications of the resonance conditions and to an extra coupling between the paraxial modes. 
The two scalar non-paraxial corrections, which are the only ones we will consider, will modify the modal phase front from $z_{\rm mode} = r^2/(2R)$ to $z_{\rm mode} \approx (1-\delta)r^2/(2R) + br^4/R^3$ with positive $\delta \ll 1$.
For typical cavity lengths not too close to concentric, $b < 0$ and the non-paraxial phase front is ``flatter" than paraboloidal.
The virtual absence of higher-order terms like $r^6$ and $r^8$ (only even orders due to rotation symmetry) limits the coupling range. 
More precisely, non-paraxial effects can only couple the fundamental $p=0$ mode to the $p=1$ and $p=2$ higher-order modes.
For a paraboloidal reference mirror, the predicted non-paraxial coupling between the $p=0$ and $p=2$ mode is
\begin{equation}
    M_{ba} = M_{ab} = - \frac{3i}{2kR} \left( \frac{R-\frac{4}{3}L}{R-L} \right) \,. 
\end{equation}
For a cavity with $R=23.8~\mu$m we calculate $M_{20}=- 0.004 \, i$ for $L=L_2=11.3~\mu$m.  
For our mirrors, which are ``steeper than paraboloidal", this extra non-paraxial contribution increases the total coupling between the $\ell=0$ modes, but only slightly.

\section{Dynamic operator $K$ and coupled modes} 

Resonances in optical cavities behave like harmonic oscillators \cite{Spreeuw1990,Haus1984}.
They have a Lorentzian spectrum under continuous excitation and the resonant field decays exponential after transient excitation.
These resonances are determined by the condition that the intra-cavity field $| \psi_i \rangle$ reproduces itself after a roundtrip, such that $M | \psi_i \rangle = \exp[i(\varphi - \varphi_i)-\gamma_i] | \psi_i \rangle$ where $\varphi = 2 k L$ and where $\varphi_i$ and $\gamma_i$ determine the resonance condition and damping of the mode. 

Below, we will demonstrate the equivalence between cavity resonances and harmonic oscillators by treating the optical cavity as a multi-transverse-mode Fabry-Perot interferometer. 
But we will start from a more general perspective.
The dynamics of any multi-mode system can be described by its dynamic operator $K$, which is related to our roundtrip operation via the relation $M \equiv \exp[i(\varphi - K)]$.
This rewrite is useful because the relevant dynamics is determined by $(1-M) \approx i(K-\varphi)$ instead of $M$ (see below) and because the expansion highlights the relevant, close-to-resonance, modes. 
The dynamic matrix can be written as $K = {\Omega} - i {\Gamma}$, with Hermitian matrices ${\Omega}$ and ${\Gamma}$, where ${\Omega}$ describes the conservative dynamics while ${\Gamma}$ describes the dissipative dynamics \cite{Spreeuw1990}. 
In the absence of losses, the roundtrip operator $M =\exp(-iK) = \exp(-i\Omega)$ is unitary (energy-conserving ) and the dynamic matrix $K$ is Hermitian. 
Just like in quantum mechanics, where the evolution operator $U = \exp(-iHt/\hbar)$ is unitary and the Hamiltonian $H$ is Hermitian. 

As an aside, we note that this general approach has among others been used to describe the dynamics of piano strings as a set of coupled harmonic oscillators \cite{Weinreich1977}.
The description of that system also involves concepts like the dynamics matrix $K$ and conservative versus dissipative coupling, which Weinreich \cite{Weinreich1977} calls reactive and resistive coupling, in analogy with electronics.

Next, we treat the optical cavity as a multi-transverse-mode Fabry-Perot interferometer. 
In a typical experiment, we illuminate the cavity with an input field $|\psi_{in} \rangle$ and observe the output field $| \psi_{out} \rangle$ as a function of cavity length or optical frequency.
The output field can be calculated by repeated operation of the roundtrip operator and summation over the infinite series of reflections to find
\begin{equation}
\label{eq:roundtrip}
    \left[ 1 - M \right] |\psi_{out} \rangle \approx i(K-\varphi) |\psi_{out} \rangle = t_1 t_2 |\psi_{in} \rangle \,.
\end{equation}
For the single-mode case, $M$ and $K$ are scalars and the solution is trivial. 
We then write $K = \varphi_i - i\gamma_i$, to find that the output field indeed varies as a complex Lorentzian, i.e. as $\propto 1/[\gamma_i+i(\varphi_i-\varphi)]$, where $\varphi_i$ is the roundtrip phase lag of mode $i$ and $(K-\varphi)$ is taken modulo $2\pi$.
The modal loss $\gamma_i = \gamma_r + \Delta \gamma_i$ combines the reflection loss $\gamma_r \approx 1 -r_1r_2$ with potential extra loss $\Delta \gamma_i$, 
The resonance will repeat itself when $\varphi = 2kL$ changes by $2\pi$. 
The ratio between this repetition and the FWHM resonance width $2 \gamma_i$ defines the modal finesse $F_i = \pi/\gamma_i$.

Next, we analyze the case for two close-to-resonance modes $a$ and $b$, which together define the field  $|\psi \rangle = \alpha |\psi_a \rangle + \beta | \psi_b \rangle$, and write Eq. (\ref{eq:roundtrip}) as
\begin{equation}
\label{eq:matrix}
    \begin{pmatrix} i(\varphi_a-\varphi) + \gamma_{a} & - M_{ab} \\ - M_{ba} & i(\varphi_b-\varphi) + \gamma_{b} \end{pmatrix} \begin{pmatrix} \alpha \\ \beta \end{pmatrix}_{out} = t_1t_2 \begin{pmatrix} \alpha \\ \beta \end{pmatrix}_{in} \,.
\end{equation}
Equation (\ref{eq:matrix}) describes the physics of two coupled harmonic oscillators.

We solve Eq. (\ref{eq:matrix}) for the case of small mixing $M_{ab}$ and dominant excitation of the less-lossy mode ($\alpha_{in} \gg \beta_{in}$ with $\gamma_a < \gamma_b$). 
The off-diagonal element will then admix a small fraction 
\begin{equation}
    \left( \frac{\beta}{\alpha} \right)_{\rm out} \approx \frac{M_{ba}}{\gamma_{ba} + i\Delta \varphi} \,,
\end{equation}
of mode $b$ into mode $a$, where $\Delta \varphi = \varphi_b - \varphi_a$ is the detuning between the resonances (= twice the difference of their single-pass Gouy phases) and $\gamma_{ba} = \gamma_b - \gamma_a$ is the excess roundtrip loss of mode $b$.
The associated change $\Delta \lambda_a$ in the eigenvalue $i \varphi_a + \gamma_a + \Delta \lambda_a$ of the mixed (dominantly $a$) mode is
\begin{equation}
    \Delta \lambda_a \approx \frac{M^2_{ba}}{\gamma_{ba}+i\Delta \varphi} \,,
\end{equation}
as $M_{ab} = M_{ba}$.
At large detuning, $|\Delta \varphi| > \gamma_{ba}$, the change $\Delta \lambda_a$ is mainly imaginary-valued.
For our case of conservative coupling, with imaginary-valued $M_{ab}=M_{ba}$, the mixing fraction $(\beta/\alpha)_{\rm out}$ is mainly real-valued and the imaginary-valued change in $\Delta \lambda_a$ results in a frequency shift of the two resonance away from each other (avoided crossing).
At smaller detuning, $\Delta \lambda_a$ also acquires a real-valued component, which corresponds to a change in loss.

\section{AFM data yield mirror profile and coupling parameter}
\begin{figure}[H]
     \centering
     \begin{subfigure}[b]{0.49\textwidth}
         \centering
         \includegraphics[width=\textwidth]{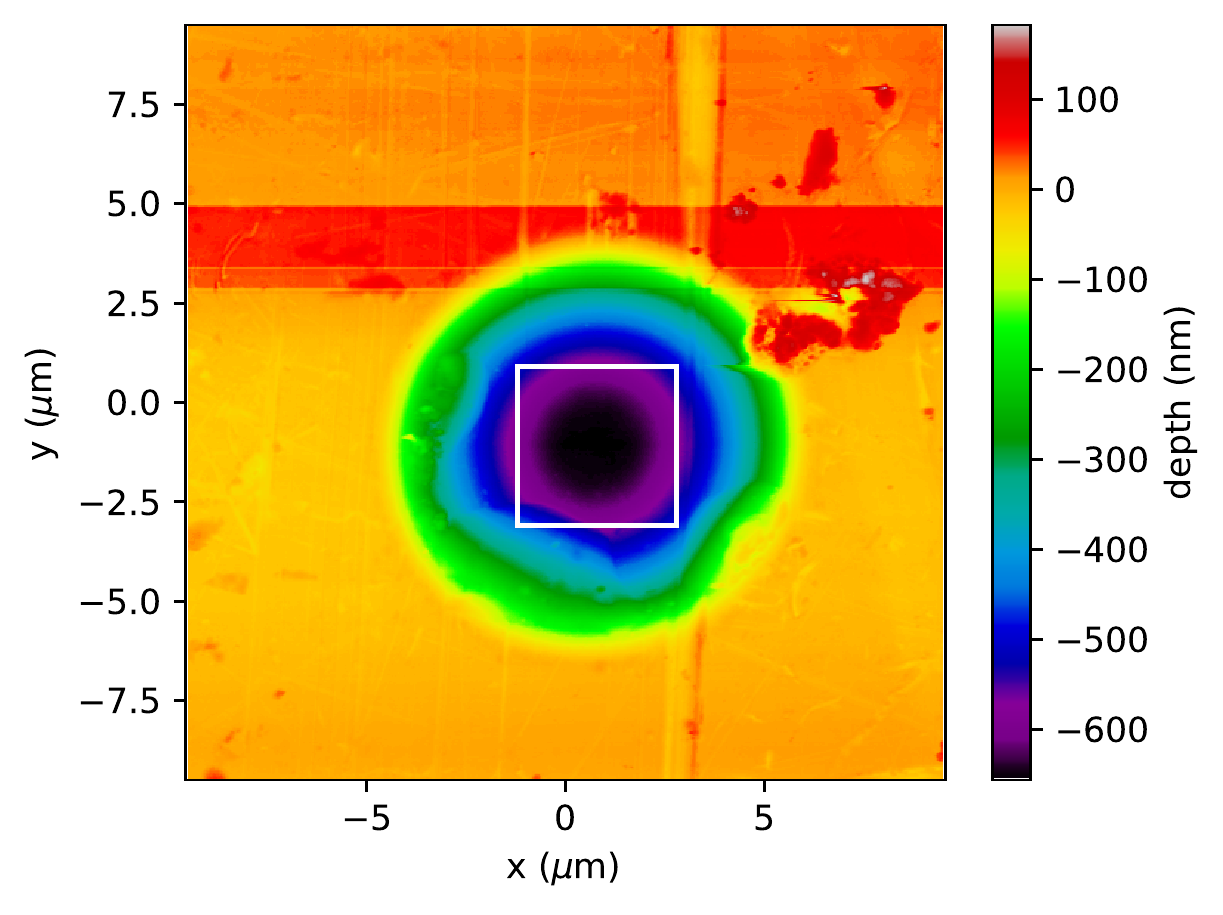}
         \caption{}
         \label{fig:AFM image full}
     \end{subfigure}
     \hfill
     \begin{subfigure}[b]{0.49\textwidth}
         \centering
         \includegraphics[width=\textwidth]{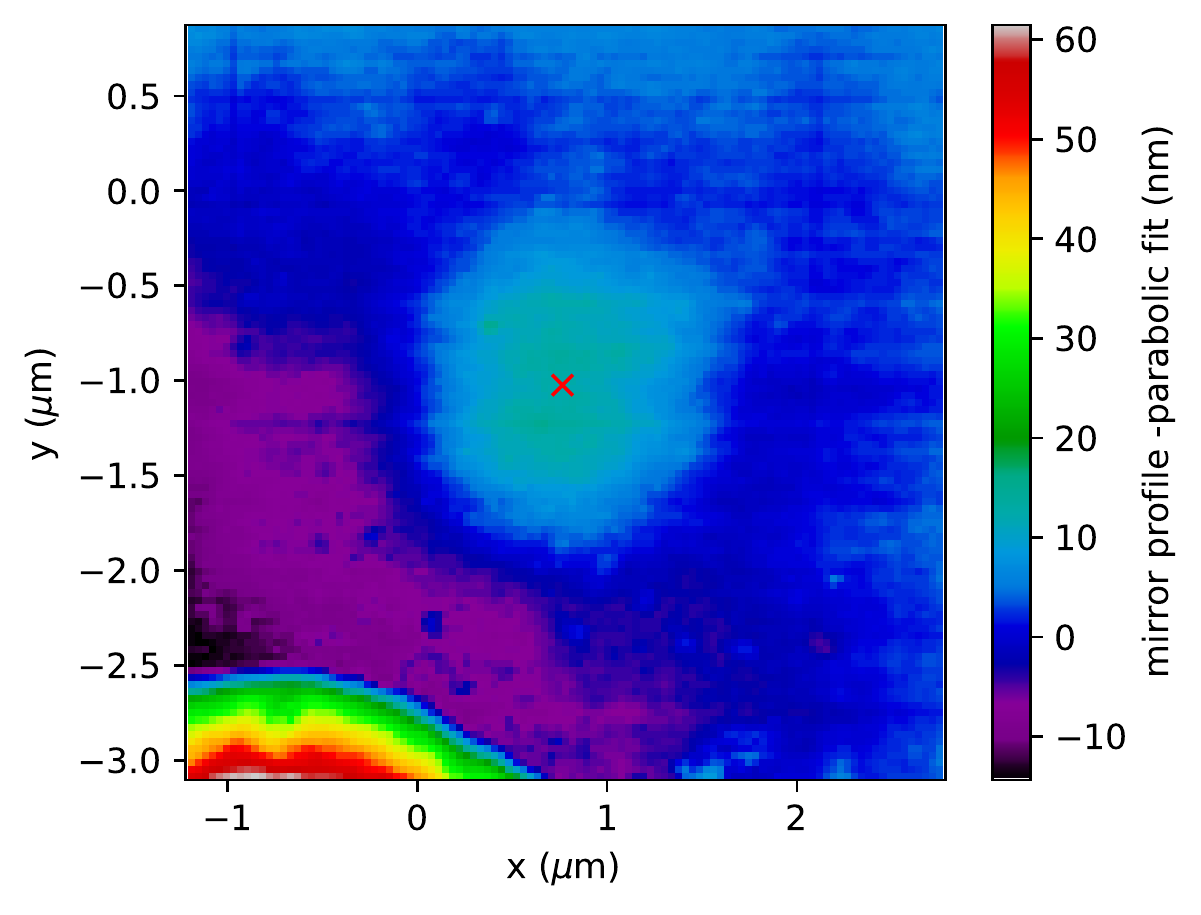}
         \caption{}
         \label{fig:AFM height difference}
     \end{subfigure}
        \caption{(a) Height profile of our concave mirror as measured with AFM. (b) Height difference $\Delta z = z - r^2/(2R)$ of mirror relative to reference paraboloid. The red cross indicates the center of this reference. }
        \label{fig:three graphs}
\end{figure}

We have measured the profile of our concave mirror with atomic force microscopy (AFM). 
Figure \ref{fig:AFM image full} shows that our mirror has a diameter $d \approx 10~\mu$m and a mirror indentation depth $D = 0.64(3)~\mu$m.
It also shows that the central region $r < 2.5~\mu$m is approximately rotationally symmetric and that deformations mainly occur for $r > 2.5~\mu$m, in particular in the lower region of the figure. These deformations are most likely due to dust and might not even affect the optical data much.

Figure \ref{fig:AFM height difference} shows the area in the white box of figure \ref{fig:AFM image full}, in an expanded view and after subtraction of a reference paraboloid $z_{\rm ref} = r^2/(2R)$ with $R=23.8~\mu$m from the height profile. 
This figure shows that the central region of the mirror sticks out, which makes the effective radius of curvature at the center larger than the reference curvature. 
It also highlights the deformations that we mentioned earlier. 

\begin{figure}[H]
    \centering
    \includegraphics[width=0.6 \linewidth]{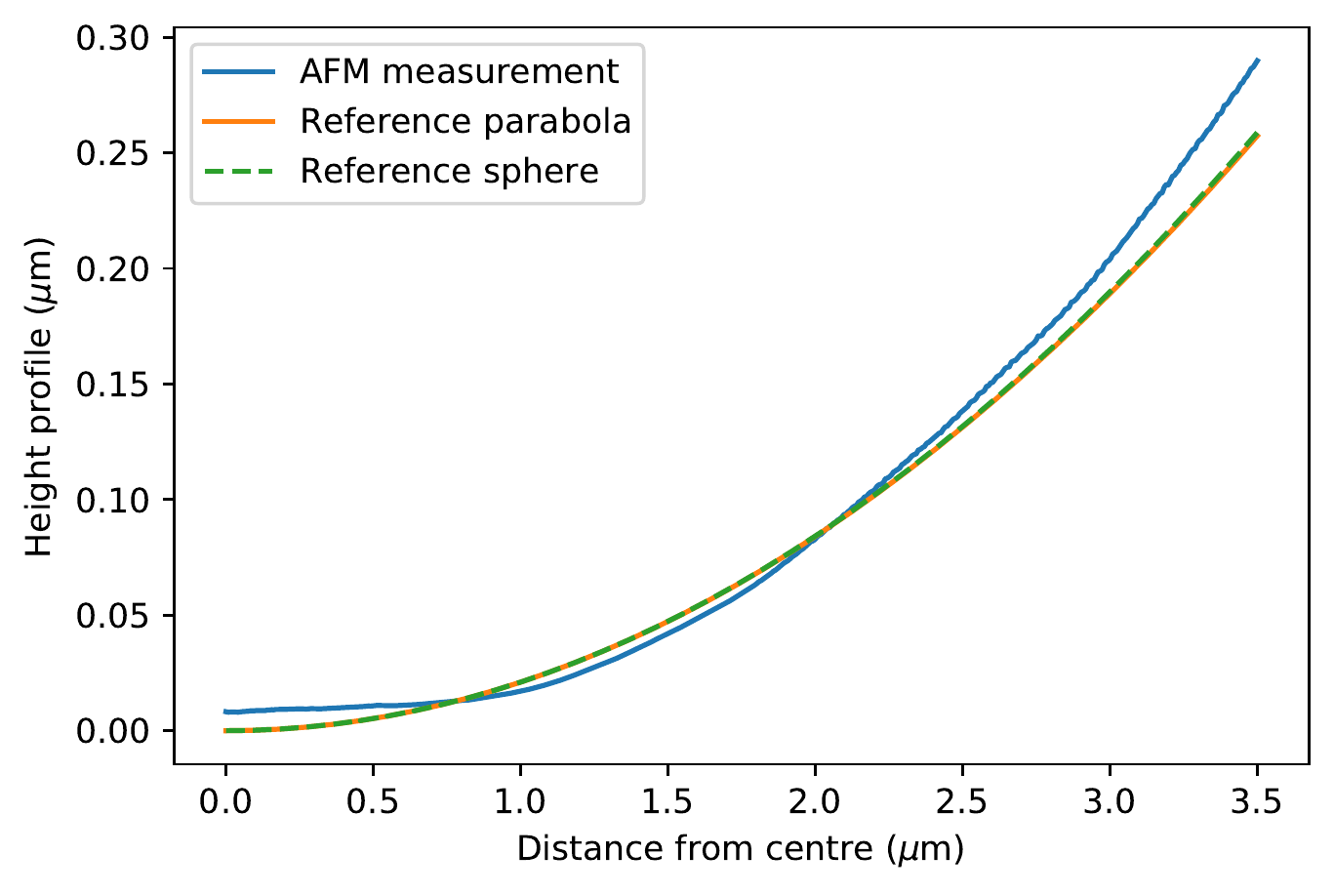}
    \caption{Rotation-averaged mirror height profile. The blue curve show the AFM measurements. The orange curve shows a reference parabola $z=r^2/(2R)$ with $R=23.8~\mu$m. The green curve shows a spherical reference surface with the same $R$.}
    \label{fig:mirror height profile}
\end{figure}

\begin{figure}[H]
    \centering
    \includegraphics[width=0.7\textwidth]{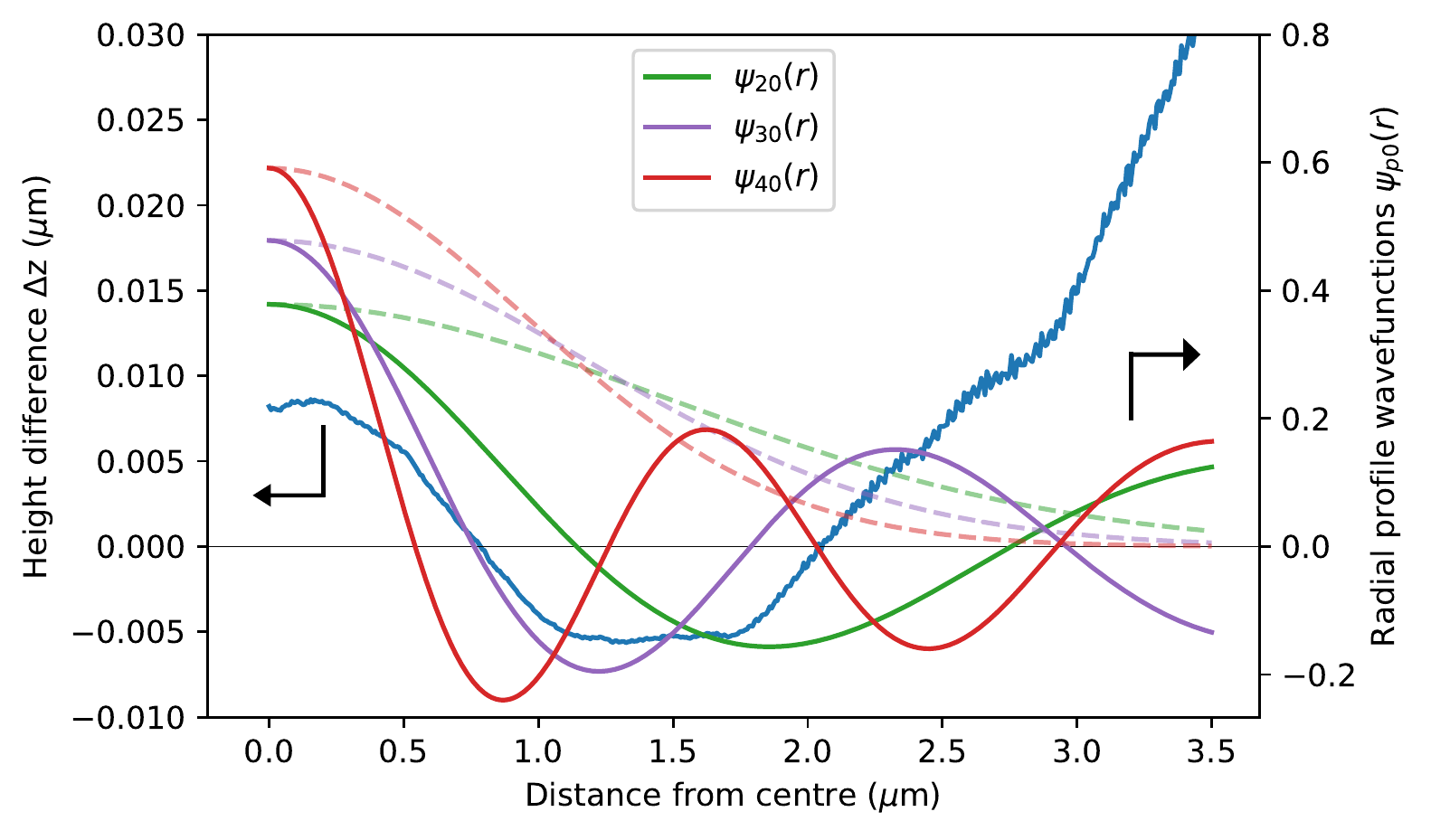}
    \caption{Height difference $\Delta z$ between the rotational-average AFM data and the ($R = 23.8~\mu$m) reference parabola. The red, purple, and green curves show the mode profile of the fundamental modes (dashed) and the nearly-degenerate higher-order ($p=4,3,$ and 2) modes around the three degenerate cavity lengths.}
    \label{fig:mirror deformation with modes}
\end{figure}

Figure \ref{fig:mirror height profile} show the rotational-average mirror height profile of the AFM data in blue.
It also shows the height profile of a reference parabola (orange) and a reference circle (green), with the same radius of curvature $R=23.8~\mu$m as deduced from the average transverse mode splittings in the optical transmission spectra (Fig. 2a in main text). 
The AFM data show that the mirror is much flatter than the reference in the center and steeper than the reference towards the edges.

Figure \ref{fig:mirror deformation with modes} compares the rotational-average height profile of the mirror with the ($R = 23.8~\mu$m) reference parabola. 
It shows (i) the height difference $\Delta z(r) = z_{\rm mirror}(r) - z_{\rm ref}(r)$ in blue and (ii) the amplitude profiles $\psi(r)$ of the fundamental mode (dashed) and the nearly-degenerate $p=4, 3$ or 2 modes (solid) in red, purple, and green. 
Note how the width $w_z$ of the fundamental mode at the concave mirror increases from $p=4$ to 3 to 2 because this width $w_z(L)$ increases monotonically with cavity length $L$.
The mode profiles are calculated at the resonance lengths $L_4 = 3.3~\mu$m, $L_3 = 5.8~\mu$m, and $L_4 = 11.3~\mu$m.

The data depicted in Fig. \ref{fig:mirror deformation with modes} allows one to calculate the coupling between the depicted LG-modes, using the equation
\begin{equation}
\label{eq:M}
    M_{p0}=(-1) \times \mel{\psi_{p0}}{-2 i k \Delta z}{\psi_{00}}= \int_0^\infty dx \,\, 2 i k \Delta z(x) \, L_p(x) \exp(-x) \,,
\end{equation}
where $x \equiv 2 r^2/w_z^2$. 
The factor $(-1)\times$ accounts for the different phase delay that the modes experience upon propagation from the flat to the curved mirror. 
By defining the height difference $\Delta z = z_{\rm mirror} - z_{\rm ref}$ with respect to a reference paraboloid, we implicitly assume that the modal phase fronts have this paraboloidal shape; this assumption proofs to be pretty good.
Inspection of Fig. \ref{fig:mirror deformation with modes} shows that the positive/negative mirror deformation $\Delta z$ generally overlaps with the positive/negative part of the higher-order mode. 
This suggests a positive coupling parameter $M_{p0}$ since the fundamental mode is positive over the full range off the mirror. 
The calculated coupling parameters are $M_{40}/i=0.015$, $M_{30}/i = 0.049$, and $M_{20}/i= 0.042$.


The analysis presented above was performed for LG-modes that are defined by the condition that their radius of curvature is $R = 23.8~\mu$m at the concave mirror.
A disadvantage of this ``fixed-radius" criterion is that the curvature of the fundamental mode is typically not perfectly matched to the curvature of the mirror. 
The fundamental matrix element $M_{00}$ is therefore not as large as it can be if that radius of curvature could be freely probed.

Alternatively, we can choose the radius of curvature of the fundamental mode based on the criterion that it should provide the best match over the modal region. 
This ``best-match" criterion translates in a two-parameter fit for the best paraboloid $z_{\rm ref}(x)=a+bx$ in the new coordinate $x \equiv 2 r^2/w_z^2$. 
This fit minimizes the rms height difference $\sigma$ defined by
\begin{equation}
\label{eq:rms}
    \sigma^2 = \mel{\psi_{00}}{\Delta z^2 }{\psi_{00}} = \int_0^\infty dx \,\,  [\Delta z(x)]^2 \, \exp(-x) \,.
\end{equation}
using linear regression.
The fit parameter $b$ yields the effective radius $R_{\rm eff}$ of the ``best-match" fundamental mode.
The calculated effective radii at the three degenerate points are $R_{\rm eff}=27.0~\mu$m for $p=4$, $R_{\rm eff}=24.8~\mu$m for $p=3$ and $R_{\rm eff}=22.7~\mu$m for $p=2$. 
This is consistent with the earlier observation that the mirror height profile is ``flatter at the center and steeper towards the edges". 
It is also consistent with the value $R = 23.8~\mu$m deduced from fits of the optical transmission spectra. 

The ``best-match" criterion yields a new beam size $w_z$ and a new set of LG-modes.
For this new set of LG modes, we again use Eq. (\ref{eq:M}) to calculate the coupling to find: $M_{40}/i= 0.015$, $M_{30}/i =0.052$ and $M_{20}/i =0.053$.
These values are different from the ones presented earlier, but not dramatically different. 
The rms fluctuations in the height profile are $\sigma \approx 4$ nm, which is significantly more than the value $\sigma \approx 2$ nm expected from the measured Finesse around 2000.
A possible explanation is that the cavity reshapes the fundamental mode in a more optimal way then we expect from the current theory.

\section{Repeated measurements on other microcavities}

We have repeated the measurements of mode coupling with other microcavities. 
We present the full analysis of one, to show that the behaviour is general.
This mirror is from a different batch than the mirror presented in the main text.

\begin{figure}[H]
    \centering
    \includegraphics[width=0.7\linewidth]{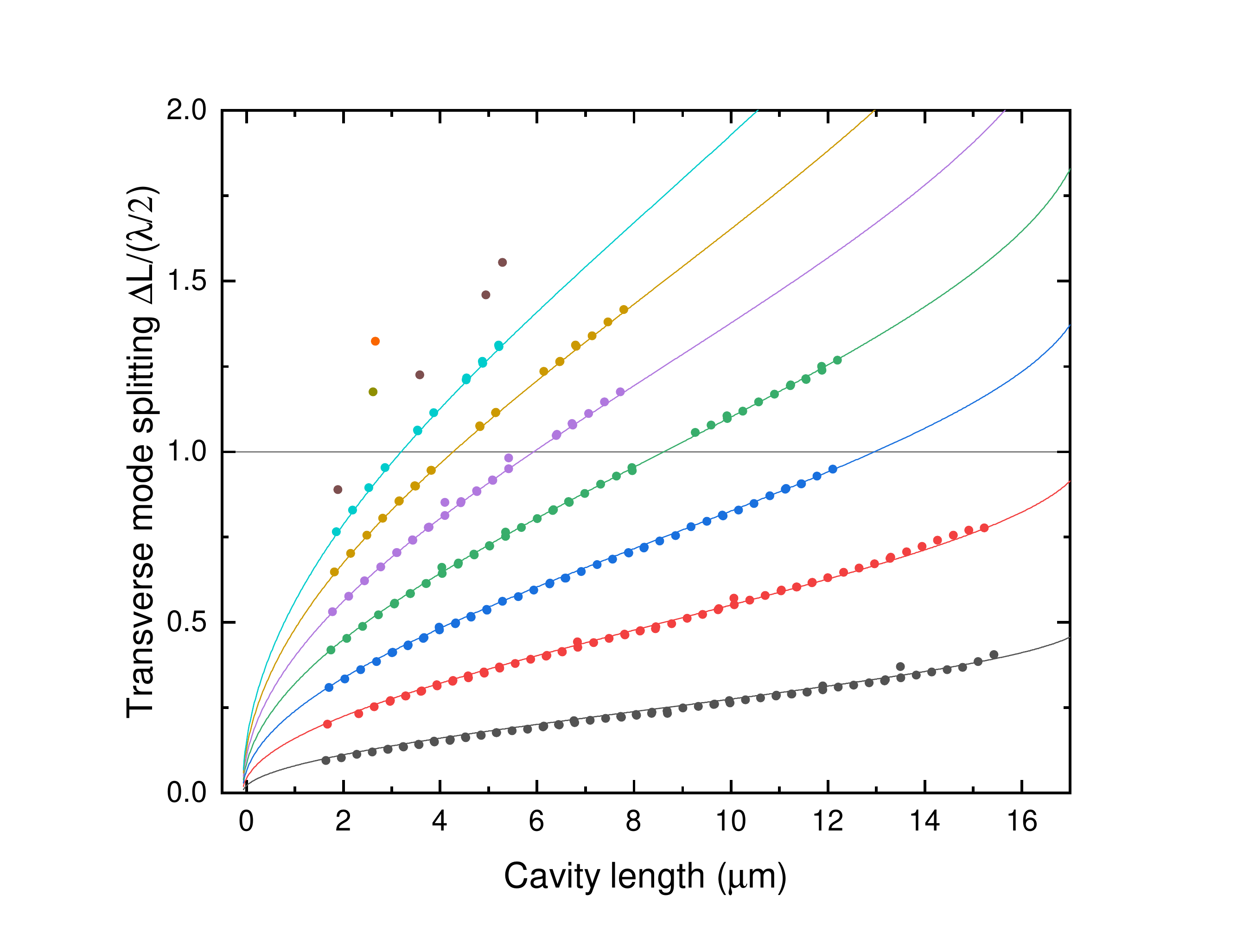}
    \caption{Transverse mode splitting versus mirror position for all transverse modes we observed.}
    \label{fig:transverse modes cavity 27april}
\end{figure}

The radius of curvature, $R$, and penetration depth, $L_D$, of the mirror are again determined from the transverse-mode spectrum.
The distance between the fundamental and higher-order transverse mode is plotted in Fig. \ref{fig:transverse modes cavity 27april}, which resembles Fig. 2a in the main text. 
The fit yields $R=17.4(1)~\mu$m and $L_D=0.04(3)~\mu$m.

The AFM measurement of this mirror (not shown) indicate that its indentation depth $D\approx0.9 ~\mu$m is larger than the value $D\approx0.6$ reported for the mirror in the main text.
The AFM data also shows that this second mirror is not as clean and has several defects as high as $0.5 ~\mu$m next to the mirror.
In practice, this means that the smallest cavity length we can reach has a longitudinal mode number $q=5$ instead of $q=3$ in the main text.
The measured finesse of this microcavity is also lower and the mirror coatings has a higher transmission.

\begin{figure}[H]
    \centering
    \includegraphics[width=0.7\linewidth]{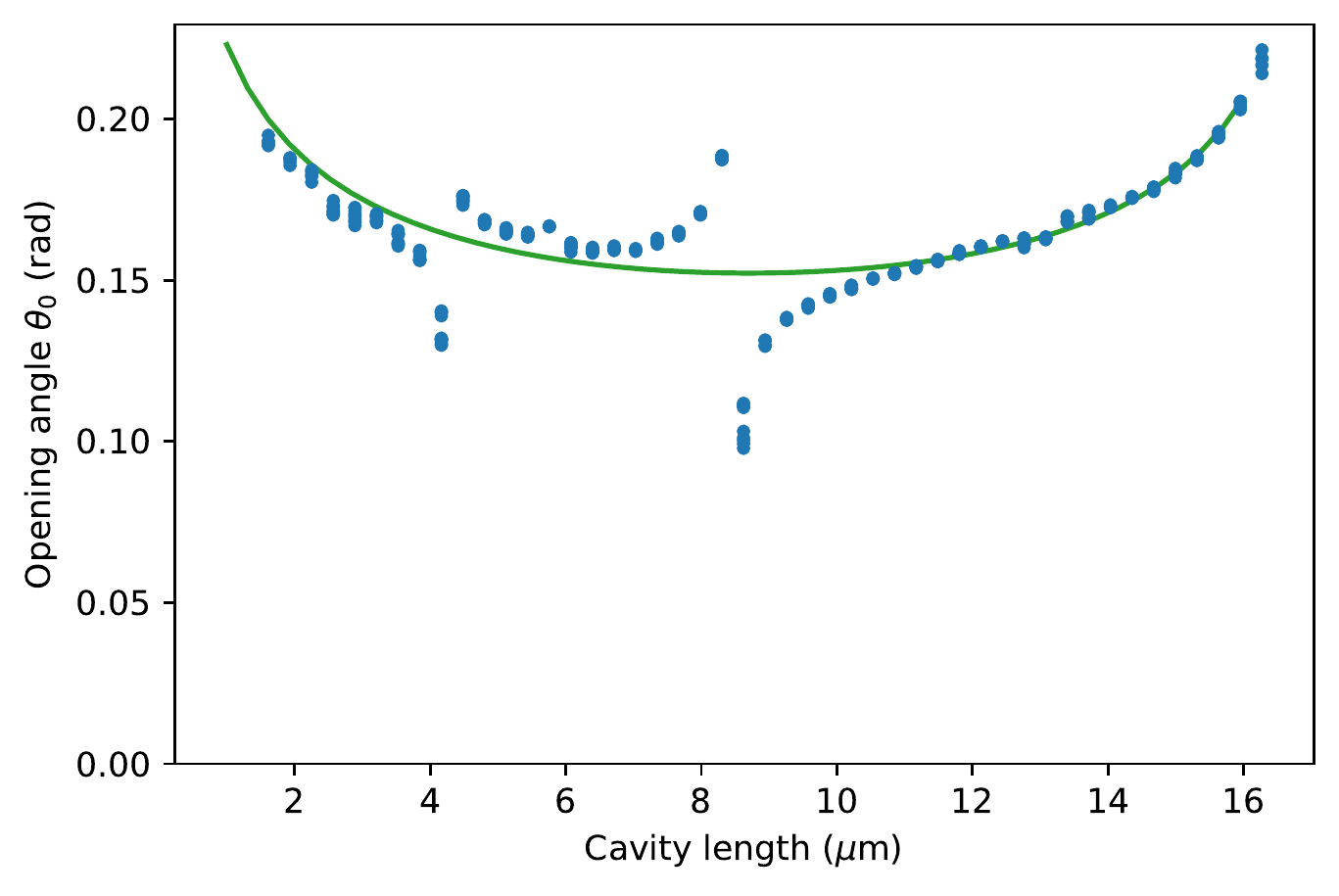}
    \caption{The Gaussian opening angle $\theta_0$ of the fundamental mode obtained from CCD images. The green line shows the theory for the uncoupled system with a radius of curvature $R=17.4~\mu$m. 
    All data points are shifted up by 6\%, to correct for a misalignment error in the imaging setup.
    }
    \label{fig:opening angle 27 april}
\end{figure}

Figure \ref{fig:opening angle 27 april} shows the measured angular width of the fundamental mode, deduced from single-Gaussian fits.
The green line shows the theoretical behaviour of the uncoupled fundamental modes;
the fitted values are increased by 6\%, to correct an error in the magnification.
The data follows the theoretical curve throughout the full cavity length, but strongly deviates at the three points of frequency degeneracy.
These deviations have the same form as the ones presented in the main text.
This means that the coupling has the same nature, and that the effect occurs for different micromirrors.
This suggests that the effect is general, rather then specific for each mirror.

\begin{figure}[H]
    \centering
    \includegraphics[width=0.7\linewidth]{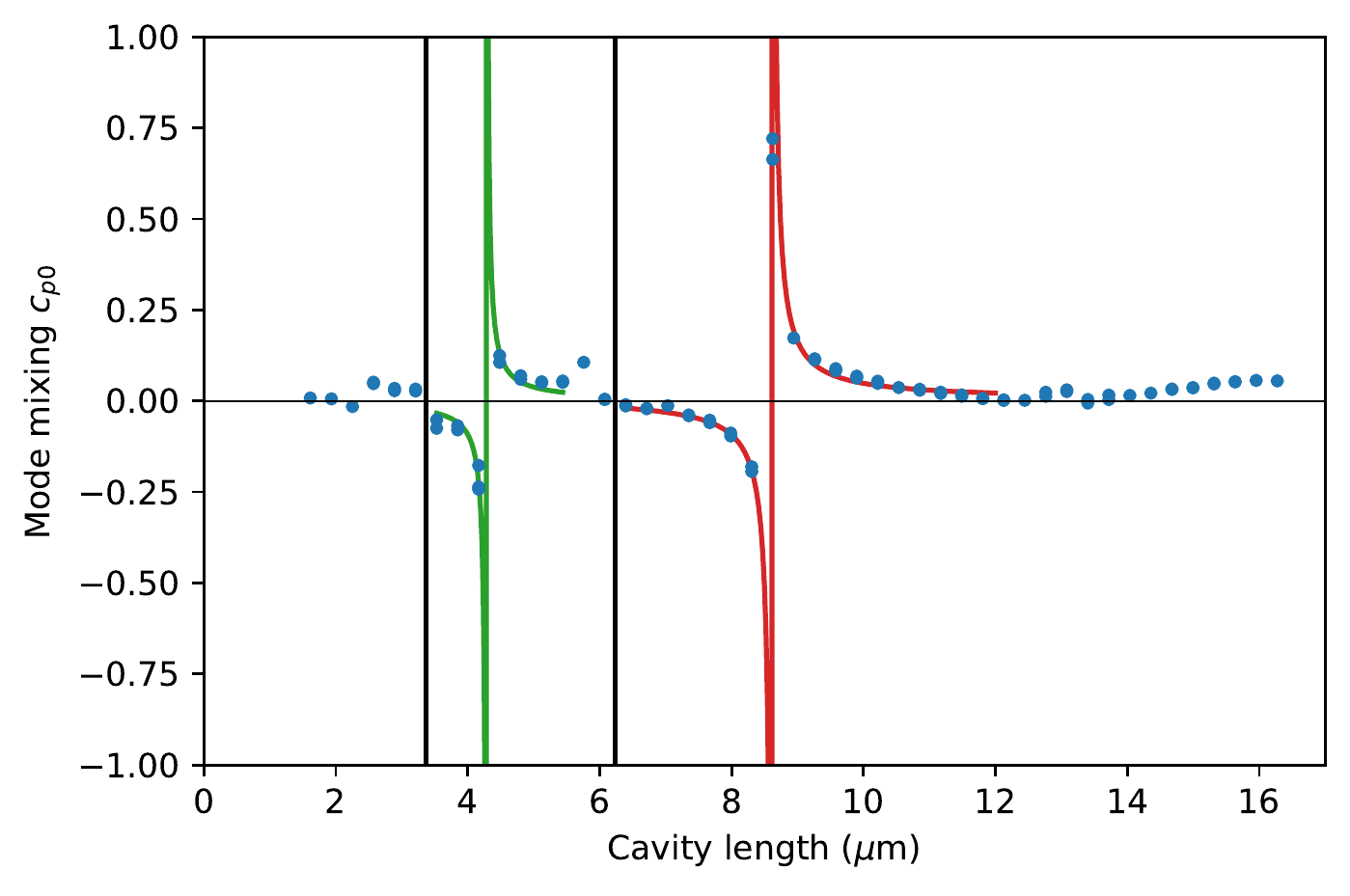}
    \caption{Mode mixing $c_{p0}=\beta/\alpha$ of modes $p=4, 3$, and $2$ into the fundamental mode. The $p=3$ (green) and $p=2$ (red) mixing is fitted to find their complex coupling parameters $M_{p0}$. The $p=4$ mixing, around $L=2.4~\mu$m is not prominent enough to allow such fitting.}
    \label{fig:coupling 27 april}
\end{figure}

Next, we use a two-mode fit for the CCD images to find the mode mixing $c_{p0}$, shown in figure \ref{fig:coupling 27 april}.
The complex coupling constant is fitted, which yield $M_{30}/i=0.021(2)$ and $M_{20}/i=0.028(1)$. 
These values are of the same order as those in the main text, where we found $M_{30}/i=0.034(3)$ and $M_{20}/i=0.029(1)$ for a different cavity.

\begin{figure}[H]
    \centering
    \includegraphics[width=0.7\linewidth]{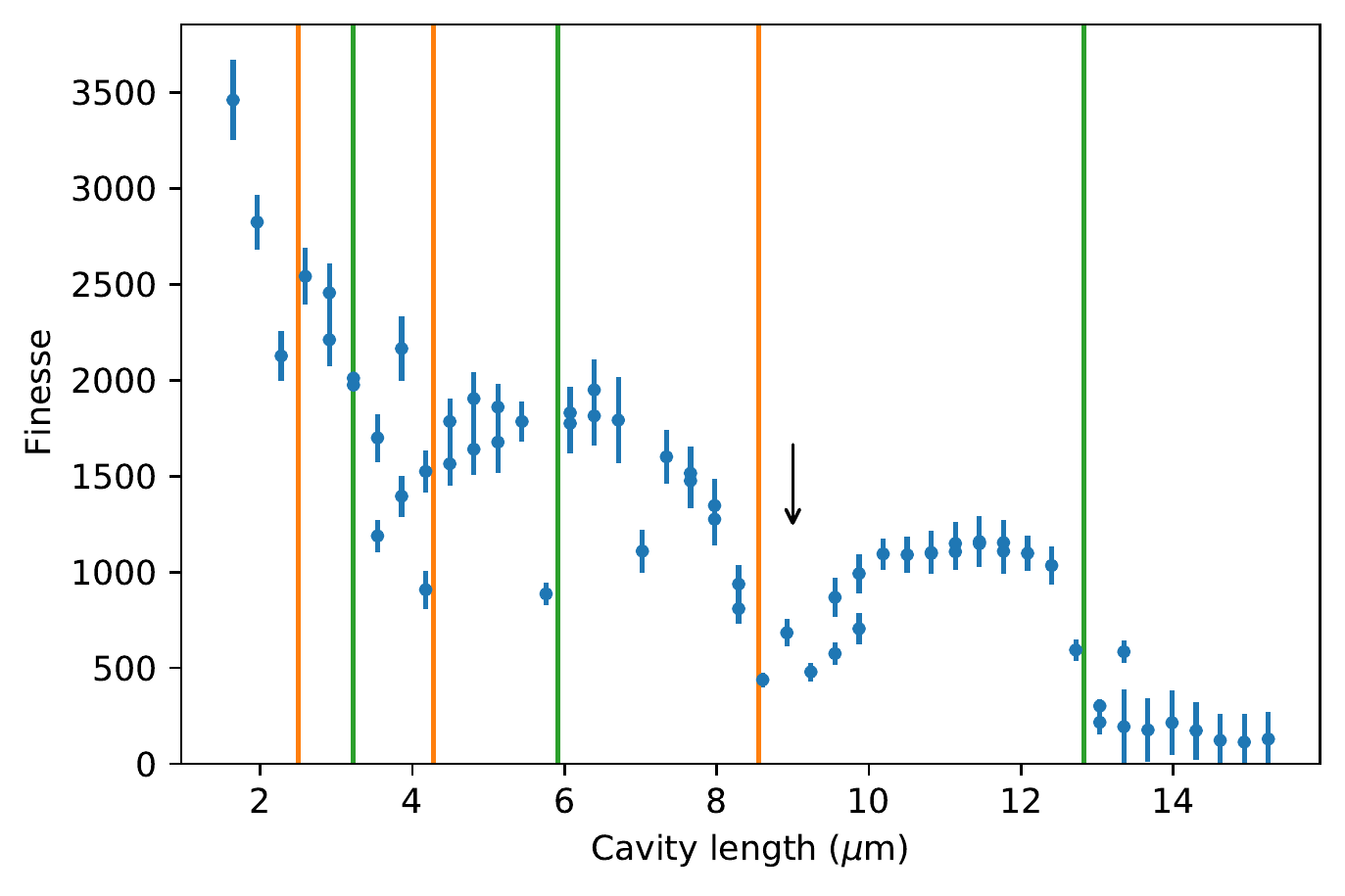}
    \caption{Finesse as a function of cavity length. The vertical lines indicate the points of frequency degeneracy, with the even modes in orange (from left to right N=8, 6, 4) and odd modes in green (N=7, 5, 3).
    }
    \label{fig:Finesse}
\end{figure}

Finally we have measured the finesse of the fundamental modes. 
Figure \ref{fig:Finesse} shows that the finesse decreases gradually with cavity length and shows prominent features around the degenerate points labeled $N=4$ and $N=3$. 
The arrow in Fig. \ref{fig:Finesse} points at a significant drop in finesse around the coupling with the $N=4$ ($p=2$) transverse modes.
The observed asymmetry of this drop is consistent with the positive sign of coupling deduced from Fig. \ref{fig:coupling 27 april}; 
at positive detuning, the size of the mode decreases at the flat mirror and increases at the curved mirror, which results in more clipping losses and a reduced finesse.
The finesses around the $N=6$ and $N=8$ degenerate points show no clear drop, indicating that mirror boundaries play no role here.
The asymmetry that we observed around the $N=3$ degenerate point is outside the scope of this paper.

\bibliography{mainsup.bbl}